\documentclass[12pt]{article}

\begin{document}

\title{\Large {\bf Hawking temperature of Kerr-Newman-AdS
                   black hole from tunneling } }

\author{{{\large Zheng Ze Ma}} \thanks{ {\sl E-mail address}:
           z.z.ma@seu.edu.cn. } $^{~,}$
          \thanks{Pen name. The author's real name is Jun Ma.}
        \\  \\
       {\normalsize {\sl Department of Physics, Southeast University,
         Nanjing, 210096, P. R. China } }}

\date{}

\maketitle

\vskip 1cm

\baselineskip 17pt

\noindent {\bf Abstract}

\vskip 10pt

  Using the null-geodesic tunneling method of Parikh and Wilczek,
we derive the Hawking temperature of a general four-dimensional
rotating black hole. In order to eliminate the motion of $\phi$
degree of freedom of a tunneling particle, we have chosen a
reference system that is co-rotating with the black hole horizon.
Then we give the explicit result for the Hawking temperature of the
Kerr-Newman-AdS black hole from the tunneling approach.

\vskip 12pt

\noindent {\sl PACS:} 04.70.Dy; 04.70.-s; 03.65.Xp

\vskip 10pt

\noindent {\sl Keywords}: Hawking temperature; Black hole; Tunneling

\vskip 1.5cm

\baselineskip 17pt

\section{Introduction}

\indent

  Since the original discovery of black hole radiation by Hawking
\cite{1}, the studies on this topic have not terminated. There are
many different methods for the derivation of Hawking radiation
\cite{2,3,4,5,6,7,8,9}. In \cite{10,11}, a semiclassical method for
the derivation of Hawking radiation was formulated by Parikh and
Wilczek based on the quantum tunneling picture. In such a method,
the radiated particles of a black hole are treated as
$s$-waves.\footnote{~ This is reasonable because for an observer at
infinity, the radiation of a black hole is spherically symmetric, no
matter whether the black hole is rotating or not.} When a particle
is radiated from the black hole horizon, it tunnels through a
barrier that is made by the tunneling particle itself due to the
horizon's contraction \cite{10,11}. To use the WKB approximation,
the tunneling rate of an $s$-wave from inside to outside the black
hole horizon is given by
$$
  \Gamma=\Gamma_{0}\exp(-2\mbox{Im}{\cal I}) ~.
  \eqno{(1)}  $$
Here, ${\cal I}$ is the action of the tunneling particle,
$\Gamma_{0}$ is a normalization factor. On the other hand, a black
hole's radiation satisfies the law of Boltzmann distribution
classically, thus the emission rate of a particle of energy $E$ from
a black hole horizon can be expressed by
$$
  \Gamma=\Gamma_{0}\exp(-\beta E) ~,
  \eqno{(2)}  $$
where $\beta=2\pi/\kappa$, $\kappa$ is the surface gravity of the
horizon. To compare (2) with (1), the Hawking temperature of a black
hole can be derived.

  After the original work of Parikh and Wilczek, many developments
on this topic have been carried out \cite{12,13,14,15}, and many
applications of this method for the derivation of Hawking radiation
of different types of black holes have been done
\cite{16,17,18,19,20,21,22,23,24,25,26,27}. In this paper, we study
the Hawking radiation of general four-dimensional rotating black
holes from the tunneling approach. We use the null-geodesic method
of Parikh and Wilczek \cite{10,11} to calculate the action of a
tunneling particle. In \cite{21}, Hawking temperature of Kerr and
Kerr-Newman black holes have been derived from tunneling approach
using dragging coordinate systems. In such a kind of coordinate
system, the spacetime of a four-dimensional rotating black hole has
been contracted to a three-dimensional slice. Thus the topology of
the spacetime of a rotating black hole has been changed to use the
method of \cite{21}. In order to keep the spacetime topology of a
rotating black hole, we choose a reference system that is
co-rotating with the event horizon to eliminate the motion of $\phi$
degree of freedom of a tunneling particle. We obtain that for a
general four-dimensional rotating black hole, its thermal radiation
temperature derived from the tunneling approach is in accordance
with its Hawking temperature derived from black hole thermodynamics.
These contents are given in Section 2. In Section 3, we give the
explicit result of the Hawking temperature of the Kerr-Newman-AdS
black hole from the tunneling approach. In Section 4, we discuss
some of the problems.

\section{Hawking temperature of four-dimensional rotating
         black holes from tunneling}

\indent

  The metric of a four-dimensional spherically symmetric black hole
can be expressed as
$$
  ds^{2}=-A(r)dt^{2}+B(r)dr^{2}+r^{2}d\Omega^{2}
  \eqno{(3)}  $$
generally. Following \cite{10,11}, the imaginary part of the action
of a tunneling particle in terms of an $s$-wave can be calculated
from
$$
  \mbox{Im}{\cal I}=\mbox{Im}\int_{r_{h}(M)}^{r_{h}(M-E)}p_{r} ~ dr
    =\mbox{Im}\int_{r_{h}(M)}^{r_{h}(M-E)}\int_{0}^{p_{r}}
     dp_{r}^{\prime} ~ dr ~,
  \eqno{(4)}  $$
where $r_{h}$ is the radius of the outer horizon, $M$ is the total
mass of the black hole, $M-E$ is the total mass of the black hole
after the particle is emitted, $E$ is the energy of the tunneling
particle. To make use of the Hamilton's equation
$$
  \dot{r}=\frac{dH}{dp_{r}}=\frac{d(M-\omega)}{dp_{r}} ~,
  \eqno{(5)}  $$
we can write
$$
  dp_{r}=\frac{d(M-\omega)}{\dot{r}} ~.
  \eqno{(6)}  $$
To substitute (6) into (4), we have
$$
  \mbox{Im}{\cal I}=\mbox{Im}\int_{r_{h}(M)}^{r_{h}(M-E)}
    \int_{0}^{E}\frac{d(M-\omega)}{\dot{r}} ~ dr =
    \mbox{Im}\int_{0}^{E}\int^{r_{h}(M)}_{r_{h}(M-E)}
    \frac{dr}{\dot{r}} ~ d\omega ~.
  \eqno{(7)}  $$
Usually, the Hawking temperature of a black hole is very small,
zero-mass particles will possess the main part of the whole
radiation spectrum. For a tunneling particle of zero-mass in terms
of an $s$-wave, it moves in a radial null geodesic. To transform the
metric (3) to the Painlev{\'e} form, $\dot{r}$ can be obtained from
$ds^{2}=0$ \cite{10,11}.

  The metric of a four-dimensional rotating black hole can be cast
in the form
$$
  ds^{2}=-g_{tt}(r,\theta)dt^{2}+g_{rr}(r,\theta)dr^{2}+
         g_{\theta\theta}(r,\theta)d\theta^{2}+
         g_{\phi\phi}(r,\theta)d\phi^{2}-2g_{t\phi}(r,\theta)dtd\phi
  \eqno{(8)}  $$
generally. For the tunneling of a rotating black hole, we can still
use the $s$-wave approximation, this is because for an observer at
infinity, the radiation of a rotating black hole is still
spherically symmetric. However, when a particle is tunneling through
the horizon of a rotating black hole, it will be dragged by the
rotation of the black hole. Thus, a tunneling particle will have
motion in the $\phi$ degree of freedom, i.e. $d\phi\neq 0$, which
means that in the calculation of the action of a tunneling particle
in formula (4), we need also to consider the contribution to the
action that comes from the motion on the $\phi$ degree of freedom,
as we can see in \cite{21}. Meanwhile, in the equation of the null
geodesic, we cannot set $d\phi=0$, thus, $\dot{r}$ cannot be
obtained from $ds^{2}=0$ conveniently.

  In order to eliminate the motion of $\phi$ degree of freedom
of a tunneling particle, we can choose a reference system that is
co-rotating with the black hole horizon. This can be realized
through the rotating coordinate transformation
$$
  \phi^{\prime}=\phi-\Omega_{h}t    ~~~~~~~~ \mbox{or} ~~~~~~~~
  \phi=\phi^{\prime}+\Omega_{h}t ~,
  \eqno{(9)}  $$
where $\Omega_{h}$ is the angular velocity of the event horizon of a
rotating black hole, which is a constant and is defined by
$$
  \Omega_{h}=\frac{g_{t\phi}}{g_{\phi\phi}}\bigg\vert_{r=r_{h}} ~.
  \eqno{(10)}  $$
In (10) and in the following, we use $r_{h}$ to represent the radius
of the event horizon of a rotating black hole. In such a co-rotating
reference system, the observers located at the horizon cannot
observe the rotation of the black hole, they will find that the
angular velocity $\Omega_{h}^{\prime}$ of the black hole is zero.
Because the tunneling of a particle takes place at the horizon, it
will not be dragged by the rotation of the black hole to observe
from such a co-rotating reference system. Therefore we have
$d\phi^{\prime}=0$ for a tunneling particle, i.e., a tunneling
particle has no motion in the $\phi^{\prime}$ degree of freedom.
This makes us be able to use equation (4) to calculate the action.
Meanwhile, in obtaining the expression of $\dot{r}$ from the
null-geodesic method, we can set $d\phi^{\prime}=0$.

  Under the coordinate transformation (9), the metric (8) turns to
the form
$$
  ds^{2}=-G_{tt}(r,\theta)dt^{2}+g_{rr}(r,\theta)dr^{2}+
         g_{\theta\theta}(r,\theta)d\theta^{2}+
         g_{\phi\phi}(r,\theta)d\phi^{\prime ~ 2}-
         2g_{t\phi}^{\prime}(r,\theta)dtd\phi^{\prime} ~,
  \eqno{(11)}  $$
where
\setcounter{equation}{11}
\begin{eqnarray}
    G_{tt} & = & g_{tt}+2g_{t\phi}\Omega_{h}-g_{\phi\phi}
      \Omega_{h}^{2} ~,           \\
  g_{t\phi}^{\prime} & = & g_{t\phi}-\Omega_{h}g_{\phi\phi} ~.
\end{eqnarray}
Because of (10), we have
$$
  g_{t\phi}^{\prime}\vert_{r=r_{h}}=0 ~.
  \eqno{(14)}  $$
This also indicates $\Omega_{h}^{\prime}=
g_{t\phi}^{\prime}/g_{\phi\phi}\vert_{r=r_{h}}=0$. On the other
hand, according to (A.5), we have
$$
  G_{tt}\big\vert_{r=r_{h}}=0 ~.
  \eqno{(15)}$$
The horizon's radius of the metric (11) is determined by
$g^{rr}\vert_{r=r_{h}}=g_{rr}^{-1}\vert_{r=r_{h}}=0$, which is the
same equation of the horizon's radius of the metric (8), thus, the
horizon's radius of a rotating black hole will not be changed under
the coordinate transformation (9). On the other hand, because of
(15), the horizon's radius for the metric (11) is also determined by
(15).

  Because in metric (11), $g_{rr}$ is singular on the horizon, in
order to calculate the action of a tunneling particle, we need to
eliminate such a coordinate singularity first. This can be realized
through the Painlev{\'e} coordinate transformation \cite{10,11}. We
use $T$ to represent the Painlev{\'e} time coordinate and make a
coordinate transformation
$$
  dt=dT-\sqrt{\frac{g_{rr}(r,\theta_{0})-1}
     {G_{tt}(r,\theta_{0})}}dr
  \eqno{(16)}$$
to the metric (11). In (16), like that in \cite{15} in studying the
tunneling from Kerr-Newman black hole, we have set $\theta$ to be a
constant in order to make the coordinate transformation integrable.
Such a manipulation is reasonable because for a tunneling particle
in terms of an $s$-wave, it satisfies $d\theta=0$, therefore we can
consider the tunneling of a particle at a constant angle
$\theta_{0}$. At last we can obtain that the physical result does
not depend on the angle $\theta_{0}$. However, the explicit integral
of (16) is not needed to be given here. Under the above coordinate
transformation, for the metric (11), we have
\setcounter{equation}{16}
\begin{eqnarray}
  ds^{2} & = & -G_{tt}(r,\theta_{0})dT^{2}+2\sqrt{G_{tt}
     (r,\theta_{0})}\sqrt{g_{rr}(r,\theta_{0})-1} ~ drdT+dr^{2}
      +g_{\phi\phi}(r,\theta_{0})d\phi^{\prime ~ 2}   \nonumber  \\
     & ~ &  -2g_{t\phi}^{\prime}(r,\theta_{0})d\phi^{\prime}
      \bigg(dT-\sqrt{\frac{g_{rr}(r,\theta_{0})-1}
      {G_{tt}(r,\theta_{0})}}dr\bigg) ~.
\end{eqnarray}
The horizon's radius for the metric (17) is determined by
$G_{tt}\big\vert_{r=r_{h}}=0$, thus, the horizon's radius for the
metric (11) is not changed after the coordinate transformation (16).
As mentioned above, for a tunneling particle in the co-rotating
reference system, it satisfies $d\phi^{\prime}=0$. Thus we have
$$
  ds^{2}=-G_{tt}(r,\theta_{0})dT^{2}+2\sqrt{G_{tt}(r,\theta_{0})}
    \sqrt{g_{rr}(r,\theta_{0})-1} ~ drdT+dr^{2} ~.
  \eqno{(18)}  $$
To suppose that the mass of the tunneling particle is zero, then its
motion is determined by the null-geodesic equation $ds^{2}=0$. To
solve this equation, we obtain
$$
  \dot{r}=\sqrt{G_{tt}(r,\theta_{0})\cdot g_{rr}(r,\theta_{0})}
    \bigg(\pm 1-\sqrt{1-\frac{1}{g_{rr}(r,\theta_{0})}} ~ \bigg) ~.
  \eqno{(19)}  $$
Because $G_{tt}\big\vert_{r=r_{h}}=0$,
$g_{rr}^{-1}\vert_{r=r_{h}}=0$, $r_{h}$ is a simple zero point of
$G_{tt}$ and $g_{rr}^{-1}$, $G_{tt}\cdot g_{rr}$ should be regular
at the horizon. The plus and minus signs in (19) correspond to
outgoing and ingoing radial null geodesics respectively. For an
outgoing tunneling particle, $\dot{r}$ is positive, we have
$$
  \dot{r}=\sqrt{G_{tt}(r,\theta_{0})\cdot g_{rr}(r,\theta_{0})}
    \bigg(1-\sqrt{1-\frac{1}{g_{rr}(r,\theta_{0})}} ~ \bigg) ~.
  \eqno{(20)}  $$

  To substitute (20) into (7), we obtain
$$
  \mbox{Im}{\cal I}=\mbox{Im}\int_{0}^{E}
      \int^{r_{h}(M)}_{r_{h}(M-E)}\frac{dr}
     {\sqrt{G_{tt}(r,\theta_{0})\cdot g_{rr}(r,\theta_{0})}
      \Big(1-\sqrt{1-\frac{1}{g_{rr}(r,\theta_{0})}}
       ~ \Big)} ~ d\omega ~.
  \eqno{(21)}  $$
To multiply $1+\sqrt{1-\frac{1}{g_{rr}(r,\theta_{0})}}$ in the
numerator and denominator of the integrand at the same time, we
obtain
$$
  \mbox{Im}{\cal I}=\mbox{Im}\int_{0}^{E}
      \int^{r_{h}(M)}_{r_{h}(M-E)}\frac
      {1+\sqrt{1-\frac{1}{g_{rr}(r,\theta_{0})}}}
      {\sqrt{G_{tt}(r,\theta_{0})\cdot g_{rr}(r,\theta_{0})} ~
      \frac{1}{g_{rr}(r,\theta_{0})}} ~ drd\omega ~.
  \eqno{(22)}  $$
For the metric of a four-dimensional rotating black hole, because
$g_{rr}$ is singular on the horizon, generally, we can write
$g_{rr}$ in the form
$$
  g_{rr}(r,\theta)=\frac{C(r,\theta)}{r-r_{h}} ~,
  \eqno{(23)}  $$
where $C(r,\theta)$ is a function regular on the horizon. To
substitute (23) into (22), we have
$$
  \mbox{Im}{\cal I}=\mbox{Im}\int_{0}^{E}
      \int^{r_{h}(M)}_{r_{h}(M-E)}\frac
      {1+\sqrt{1-\frac{r-r_{h}}{C(r,\theta_{0})}}}
      {\sqrt{G_{tt}(r,\theta_{0})\cdot g_{rr}(r,\theta_{0})} ~
      \frac{r-r_{h}}{C(r,\theta_{0})}} ~ drd\omega ~.
  \eqno{(24)}  $$
In (24), $r_{h}$ is a simple pole of the integrand. To add a small
imaginary part to the variable $r$, and to let the integral path
round the pole in a semicircle, the integral of $dr$ can be
evaluated which results
$$
  \mbox{Im}{\cal I}=2\pi\int_{0}^{E}\frac{C(r_{h},\theta_{0})}
    {\sqrt{G_{tt}(r_{h},\theta_{0})\cdot g_{rr}(r_{h},\theta_{0})}}
     ~ d\omega ~.
  \eqno{(25)}  $$
It is reasonable to suppose that the energy $E$ of the tunneling
particle is far less than the total mass $M$ of the black hole, i.e.
$E<<M$, thus, in (25), the integrand can be treated as a constant.
Therefore we obtain
$$
  \mbox{Im}{\cal I}=2\pi E\frac{C(r_{h},\theta_{0})
    \sqrt{g^{rr}(r_{h},\theta_{0})}}
    {\sqrt{G_{tt}(r_{h},\theta_{0})}} ~.
  \eqno{(26)}  $$

  Because $G_{tt}(r_{h},\theta)=0$, $g^{rr}(r_{h},\theta)=0$, near
the horizon, we can expand $G_{tt}(r,\theta_{0})$ and
$g^{rr}(r,\theta_{0})$ in the form
$$
  G_{tt}(r,\theta_{0})=G_{tt}^{\prime}(r_{h},\theta_{0})
    (r-r_{h})+\ldots ~,
  \eqno{(27)}  $$
$$
  g^{rr}(r,\theta_{0})=g^{rr\prime}(r_{h},\theta_{0})
    (r-r_{h})+\ldots ~,
  \eqno{(28)}  $$
where in (27) and (28), $\ldots$ represents high order terms of
$(r-r_{h})$. From (23), we have
$$
  g^{rr\prime}(r_{h},\theta_{0})=\frac{1}{C(r_{h},\theta_{0})} ~.
  \eqno{(29)}  $$
To substitute (27)--(29) into (26), we obtain
$$
  \mbox{Im}{\cal I}=\frac{2\pi E}{\sqrt{G_{tt}^{\prime}
     (r_{h},\theta_{0})g^{rr\prime}(r_{h},\theta_{0})}} ~.
  \eqno{(30)}  $$
To substitute (30) into (1), we can see that the tunneling rate can
be cast in the form of (2), which is the Boltzmann distribution, and
we obtain
$$
  \mbox{Im}{\cal I}=\frac{\pi E}{\kappa(r_{h})} ~.
  \eqno{(31)}  $$
To compare (31) with (30), we obtain
$$
  \kappa(r_{h})=\frac{\sqrt{G_{tt}^{\prime}(r_{h},\theta_{0})
     g^{rr\prime}(r_{h},\theta_{0})}}{2} ~.
  \eqno{(32)}  $$
Thus, we obtain the thermal temperature of a four-dimensional
rotating black hole
$$
  T_{H}=\frac{\sqrt{G_{tt}^{\prime}(r_{h},\theta_{0})
     g^{rr\prime}(r_{h},\theta_{0})}}{4\pi} ~.
  \eqno{(33)}  $$
Equation (33) is derived from the tunneling approach. On the other
hand, in Appendix A, we have derived a formula (A.12) for the
surface gravity of a four-dimensional rotating black hole from black
hole thermodynamics which is given by
$$
  \kappa(r_{h})=\lim_{r\rightarrow r_{h}}
     \frac{\partial_{r}\sqrt{G_{tt}}}{\sqrt{g_{rr}}}=
     \lim_{r\rightarrow r_{h}}
     \frac{\partial_{r}G_{tt}}{2\sqrt{G_{tt}\cdot g_{rr}}} ~.
  \eqno{(34)}$$
From black hole thermodynamics \cite{28,29}, we know that on the
horizon, $\kappa(r_{h})$ is a constant, therefore we can evaluate it
at an arbitrary angle $\theta_{0}$. To substitute (27) and (28) into
(34), we obtain
$$
  \kappa(r_{h})=\frac{\sqrt{G_{tt}^{\prime}(r_{h},\theta_{0})
     g^{rr\prime}(r_{h},\theta_{0})}}{2} ~.
  \eqno{(35)}  $$
To compare (32) with (35), we can see that they are equivalent.
Because $\kappa(r_{h})$ is a constant on the horizon, the explicit
result for the surface gravity of a rotating black hole obtained
from (35) will not depend on the parameter $\theta_{0}$. This means
that in (32) and (33), the explicit results for the surface gravity
and Hawking temperature of a rotating black hole will not depend on
the parameter $\theta_{0}$ either.

\section{Hawking temperature of Kerr-Newman-AdS black hole}

\indent

  In this section, we derive the Hawking temperature of the
Kerr-Newman-AdS black hole using (33) of Section 2. In the
Boyer-Lindquist coordinates, the metric of the Kerr-Newman-AdS is
given by \cite{30}
\setcounter{equation}{35}
\begin{eqnarray}
  ds^{2}= & - & \!\! \frac{1}{\Sigma}[\Delta_{r}-
     \Delta_{\theta}a^{2}\sin^{2}\theta]dt^{2}+
     \frac{\Sigma}{\Delta_{r}}dr^{2}+
     \frac{\Sigma}{\Delta_{\theta}}d\theta^{2}+
     \frac{1}{\Sigma\Xi^{2}}[\Delta_{\theta}(r^{2}+a^{2})^{2}
     \nonumber       \\
    & ~ & -\Delta_{r}a^{2}\sin^{2}\theta]\sin^{2}\theta d\phi^{2}
    -\frac{2a}{\Sigma\Xi}[\Delta_{\theta}(r^{2}+a^{2})-
     \Delta_{r}]\sin^{2}\theta dtd\phi ~,
\end{eqnarray}
where
$$
  \Sigma=r^{2}+a^{2}\cos^{2}\theta ~,   ~~~~~~
  \Xi=1+\frac{1}{3}\Lambda a^{2} ~,
  \eqno{(37)}  $$
$$
  \Delta_{\theta}=1+\frac{1}{3}\Lambda a^{2}\cos^{2}\theta ~,
   ~~~~~~ \Delta_{r}=(r^{2}+a^{2})
  \big(1-\frac{1}{3}\Lambda r^{2}\big)-2Mr+Q^{2} ~,
  \eqno{(38)}  $$
$\Lambda$ is the cosmological constant, $\Lambda<0$. The horizons of
the metric (36) are determined by
\setcounter{equation}{38}
\begin{eqnarray}
  \Delta_{r} & = & (r^{2}+a^{2})\big(1-\frac{1}{3}
     \Lambda r^{2}\big)-2Mr+Q^{2}     \nonumber     \\
     & = & -\frac{1}{3}\Lambda\bigg[r^{4}-\Big(\frac{3}{\Lambda}
       -a^{2}\Big)r^{2}+\frac{6M}{\Lambda}r-\frac{3}{\Lambda}
       (a^{2}+Q^{2})\bigg]     \nonumber     \\
     & = & -\frac{1}{3}\Lambda(r-r_{++})(r-r_{--})
     (r-r_{+})(r-r_{-})=0 ~.
\end{eqnarray}
The equation $\Delta_{r}=0$ has four roots \cite{31}, where $r_{++}$
and $r_{--}$ are a pair of complex conjugate roots, $r_{+}$ and
$r_{-}$ are two real positive roots, and we suppose $r_{+}>r_{-}$.
Thus, $r=r_{+}$ is the event horizon. We first calculate the Hawking
temperature at a special value of $\theta_{0}$ and we choose
$\theta_{0}=0$. To expand $g^{rr}(r,\theta_{0}=0)$ near the event
horizon $r_{h}=r_{+}$, we obtain
$$
  g^{rr}(r,\theta_{0}=0)=\frac{-\frac{1}{3}\Lambda(r_{+}-r_{++})
    (r_{+}-r_{--})(r_{+}-r_{-})}{r_{+}^{2}+a^{2}}(r-r_{+})+\ldots ~,
  \eqno{(40)}  $$
where $\ldots$ are high order terms of $(r-r_{+})$. $G_{tt}$ is
defined by (12). We can rewrite it in the form
$$
  G_{tt}=g_{tt}+g_{t\phi}\Omega_{h}+
    (g_{t\phi}-\Omega_{h}g_{\phi\phi})\Omega_{h}
    =g_{tt}+g_{t\phi}\Omega_{h}+g_{t\phi}^{\prime}\Omega_{h}
  \eqno{(41)}  $$
generally. According to (14), $g_{t\phi}^{\prime}$ is zero on the
horizon, thus the last term of (41) does not need to be considered
when we expand $G_{tt}(r,\theta_{0})$ near the horizon. The angular
velocity of the Kerr-Newman-AdS black hole defined by (10) is
$\Omega_{h}=\frac{a\Xi}{r_{+}^{2}+a^{2}}$. At $\theta_{0}=0$,
$G_{tt}(r,\theta_{0}=0)$ can be expanded as
$$
  G_{tt}(r,\theta_{0}=0)=\frac{-\frac{1}{3}\Lambda(r_{+}-r_{++})
    (r_{+}-r_{--})(r_{+}-r_{-})}{r_{+}^{2}+a^{2}}(r-r_{+})+\ldots ~,
  \eqno{(42)}  $$
where $\ldots$ are high order terms of $(r-r_{+})$. To compare (42)
and (40) with (27) and (28), we can obtain
$G_{tt}^{\prime}(r_{+},\theta_{0}=0)$ and
$g^{rr\prime}(r_{+},\theta_{0}=0)$. To substitute
$G_{tt}^{\prime}(r_{+},\theta_{0}=0)$ and
$g^{rr\prime}(r_{+},\theta_{0}=0)$ into (33), we obtain, for the
Kerr-Newman-AdS black hole,
$$
  T_{H}=-\frac{\Lambda}{12\pi(r_{+}^{2}+a^{2})}(r_{+}-r_{++})
      (r_{+}-r_{--})(r_{+}-r_{-}) ~.
  \eqno{(43)}  $$
Because $\Lambda<0$, $r_{+}$ and $r_{-}$ are positive,
$r_{+}>r_{-}$, $r_{++}$ and $r_{--}$ are complex conjugate, these
make sure that $T_{H}$ is positive. At an arbitrary value of
$\theta_{0}$, through explicit calculation, $g^{rr}(r,\theta_{0})$
and $G_{tt}(r,\theta_{0})$ can be expanded as
\setcounter{equation}{43}
\begin{eqnarray}
  g^{rr}(r,\theta_{0}) & = & \frac{-\frac{1}{3}\Lambda
    (r_{+}-r_{++})(r_{+}-r_{--})(r_{+}-r_{-})}{r_{+}^{2}+a^{2}
      \cos^{2}\theta_{0}}(r-r_{+})+\ldots ~,                \\
  G_{tt}(r,\theta_{0}) & = & \frac{-\frac{1}{3}
    \Lambda(r_{+}-r_{++})(r_{+}-r_{--})
    (r_{+}-r_{-})(r_{+}^{2}+a^{2}\cos^{2}\theta_{0})}
    {(r_{+}^{2}+a^{2})^{2}}(r-r_{+})+\ldots ~.
\end{eqnarray}
To compare (45) and (44) with (27) and (28), we can
obtain $G_{tt}^{\prime}(r_{+},\theta_{0})$ and
$g^{rr\prime}(r_{+},\theta_{0})$. To substitute
$G_{tt}^{\prime}(r_{+},\theta_{0})$ and
$g^{rr\prime}(r_{+},\theta_{0})$ into (33), we obtain again
$$
  T_{H}=-\frac{\Lambda}{12\pi(r_{+}^{2}+a^{2})}(r_{+}-r_{++})
      (r_{+}-r_{--})(r_{+}-r_{-}) ~.
  \eqno{(46)}  $$
In \cite{32}, another expression for the Hawking temperature of the
Kerr-Newman-AdS black hole has been obtained which is given by
$$
  T_{H}=\frac{3r_{+}^{4}+(a^{2}+l^{2})r_{+}^{2}-l^{2}(a^{2}+Q^{2})}
       {4\pi l^{2}r_{+}(r_{+}^{2}+a^{2})} ~,
  \eqno{(47)}  $$
where $\Lambda=-3/l^{2}$. It is not difficult to verify that these
two expressions of $T_{H}$ for the Kerr-Newman-AdS black hole are
equivalent. The result of (46) is also equal to that obtained from
(A.13) and (A.14). From this example, we can also see that the
explicit result of the Hawking temperature given by (33) does not
depend on the parameter $\theta_{0}$.

  In the case $\Lambda=0$, the metric (36) degenerates to the metric
of four-dimensional Kerr-Newman black hole. If the charge is zero,
the metric will be the Kerr black hole. Following the same approach
as above, we can also obtain their Hawking temperature from
tunneling.

\section{Discussion}

\indent

  In this paper, we have studied the Hawking radiation of general
four-dimensional rotating black holes using the tunneling method of
Parikh and Wilczek \cite{10,11}. We obtain that the tunneling rate
of a zero-mass particle is given by
$$
  \Gamma=\Gamma_{0}\exp(-\beta E)=\Gamma_{0}\exp(-E/T_{H}) ~,
  \eqno{(48)}  $$
which is just the Boltzmann distribution. The thermal temperature
$T_{H}$ of a four-dimensional rotating black hole is given by (33),
which is in accordance with the Hawking temperature derived from
black hole thermodynamics. And we have given the explicit result for
the Hawking temperature of the Kerr-Newman-AdS black hole from the
tunneling approach. In order to eliminate the motion of $\phi$
degree of freedom of a tunneling particle from a rotating black
hole, we choose a reference system that is co-rotating with the
black hole horizon. In such a co-rotating reference system, we
avoided the dimension degeneration in the method of dragging
coordinate system adopted in \cite{21} for the tunneling of a
rotating black hole.

  It is necessary to point out that if we use the method of \cite{21}
to calculate the action of a tunneling particle directly for the
Kerr-Newman-AdS black hole, then we need to consider the action that
comes from the motion on the $\phi$ degree of freedom, the
calculation will be rather complicated in this case. In order to
simplify the calculation, we have made a rotating coordinate
transformation first. At the same time, the method provided in this
paper is general for a general four-dimensional rotating black hole.
And then we applied our result to the special case of the
Kerr-Newman-AdS black hole. Another point needed to point out here
is that there are some overlaps between the approach of this paper
and the manipulation of the tunneling from the Kerr-Newman black
hole in \cite{15} using the null-geodesic method. The difference
lies in that in \cite{15} the rotating coordinate transformation for
the tunneling of a rotating black hole was not proposed clearly, and
it has not been used to a general four-dimensional rotating black
hole. While in this paper, we have studied the Hawking temperature
of a general four-dimensional rotating black hole from tunneling
using the rotating coordinate transformation clearly, and then we
applied our result to the special case of the Kerr-Newman-AdS black
hole. An alternative method for the calculation of the action of a
tunneling particle was proposed in \cite{14} from the
Hamilton--Jacobi equation approach. Such a method was applied to the
tunneling of some rotating black holes in \cite{14,15}. For the
tunneling of the Kerr-Newman-AdS black hole, to use the
Hamilton--Jacobi equation method of \cite{14}, we can also make a
rotating coordinate transformation first to simplify the
calculation. The same results of (33) and (46) will be obtained at
last. However, limited by the length of this paper, we will not give
such a derivation further in this Letter.

  The tunneling rate (48) and Hawking temperature (33) for a rotating
black hole are obtained in the reference system co-rotating with the
black hole horizon. However, because the obtained tunneling rate and
Hawking temperature of a black hole are scalars, they will not
change for an observer static relatively to infinity. Thus, we can
deduce that for an observer static relatively to infinity, the
tunneling rate and Hawking temperature of a four-dimensional
rotating black hole are still given by (48) and (33). The difference
lies in that, for a tunneling particle, or an observer, the angular
velocity of a rotating black hole is zero in the co-rotating
reference system, while it is $\Omega_{h}$ in the static reference
system. To combine the first law of black hole thermodynamics, we
can generalize the tunneling rate (48) to a particle with non-zero
angular momentum and non-zero charge.

\vskip 1cm

\noindent{\bf \Large Acknowledgements}

\indent

  The author is grateful very much for the comments made by the
referee which have improved the contents of this paper.

\vskip 2cm

\noindent{\bf \Large Appendix A. Hawking temperature of
four-dimensional rotating black holes from black hole
thermodynamics}

\indent

  In this appendix, we give an expression for the Hawking temperature
of a four-dimensional rotating black hole from black hole
thermodynamics. The metric of a four-dimensional rotating black hole
is given by (8) generally. For the metric (8), there exists the
Killing field
$$
  \xi^{\mu}=\frac{\partial}{\partial t}+
       \Omega_{h}\frac{\partial}{\partial \phi} ~,
  \eqno{({\rm A}.1)}$$
where $\Omega_{h}$ is the angular velocity of the horizon which is a
constant. Here, we mean that the horizon is the outer horizon for a
rotating black hole. Because the horizon is a null surface and
$\xi^{\mu}$ is normal to the horizon, we have on the horizon
\cite{28}
$$
  \xi^{\mu}\xi_{\mu}\vert_{r=r_{h}}=0 ~.
  \eqno{({\rm A}.2)}$$
For the metric (8), we have
$$
  \xi^{\mu}\xi_{\mu}=g_{tt}+2g_{t\phi}\Omega_{h}
     -g_{\phi\phi}\Omega_{h}^{2} ~.
  \eqno{({\rm A}.3)}$$
Here, we have defined that the square of the norm of the Killing
field is positive outside the horizon, at least for the case
$\Omega_{h}=0$. As in (12), we define
$$
    G_{tt}=g_{tt}+2g_{t\phi}\Omega_{h}-g_{\phi\phi}
      \Omega_{h}^{2} ~.
  \eqno{({\rm A}.4)}$$
Thus we have
$$
    G_{tt}\big\vert_{r=r_{h}}=0 ~.
  \eqno{({\rm A}.5)}$$
Following \cite{28} we write
$$
  \xi^{\mu}\xi_{\mu}=-\lambda^{2} ~,
  \eqno{({\rm A}.6)}$$
where $\lambda$ is a scalar function, and it is a constant on the
horizon. According to (A.3), we have $\lambda^{2}=-G_{tt}$ for the
metric (8) of a four-dimensional rotating black hole. Let
$\nabla^{\mu}$ represent the covariant derivative operator, thus
$\nabla^{\mu}(\xi^{\nu}\xi_{\nu})$ is also normal to the horizon.
Then, according to \cite{28,29}, there exists a function $\kappa$
satisfying the equation
$$
  \nabla^{\mu}(-\lambda^{2})=-2\kappa\xi^{\mu} ~,
  \eqno{({\rm A}.7)}$$
where on the horizon $\kappa(r_{h})$ is a constant and is just the
horizon's surface gravity.

  Similarly, we have the lower index equation
$$
  \nabla_{\mu}(-\lambda^{2})=-2\kappa\xi_{\mu} ~.
  \eqno{({\rm A}.8)}$$
Thus, from (A.7) and (A.8) we have
$$
  \nabla^{\mu}(-\lambda^{2})\nabla_{\mu}(-\lambda^{2})
   =-4\kappa^{2}\lambda^{2} ~.
  \eqno{({\rm A}.9)}$$
Because $\lambda^{2}$ is a scalar function, $\kappa^{2}$ is also a
scalar function. Therefore the surface gravity of a black hole
horizon is invariant under general coordinate transformations,
including the rotation of (9). From (A.3), (A.4), (A.6), (A.9), and
the axial symmetry of the metric, we obtain
$$
  4\kappa^{2}G_{tt}=g^{rr}(\partial_{r}G_{tt})^{2}
         + g^{\theta\theta}(\partial_{\theta}G_{tt})^{2} ~.
  \eqno{({\rm A}.10)}$$
Because of (A.5), we have
$$
  \lim_{r\rightarrow r_{h}}\partial_{\theta}G_{tt}=0 ~.
  \eqno{({\rm A}.11)}$$
Therefore, to take the limit $r\rightarrow r_{h}$ in both sides of
(A.10) yields
$$
  \kappa(r_{h})=\lim_{r\rightarrow r_{h}}
     \frac{\partial_{r}\sqrt{G_{tt}}}{\sqrt{g_{rr}}} ~.
  \eqno{({\rm A}.12)}$$
In (A.12), because $G_{tt}$ is zero on the horizon, the partial
derivative is taken before the limit. Thus, the Hawking temperature
of the metric (8) is given by
$$
  T_{H}=\frac{\kappa(r_{h})}{2\pi}=\lim_{r\rightarrow r_{h}}
     \frac{\partial_{r}\sqrt{G_{tt}}}{2\pi\sqrt{g_{rr}}} ~.
  \eqno{({\rm A}.13)}$$
Because $\kappa(r_{h})$ is a constant on the horizon \cite{28,29},
it can be evaluated at an arbitrary $\theta$. For convenience, it
can be evaluated at $\theta=0$ usually. For the metrics of many
four-dimensional rotating black holes, we can see that usually they
satisfy $G_{tt}\vert_{\theta=0}=g_{tt}\vert_{\theta=0}$. Thus we can
write
$$
  T_{H}=\frac{\kappa(r_{h})}{2\pi}=
     \lim_{\theta=0, ~ r\rightarrow r_{h}}
     \frac{\partial_{r}\sqrt{g_{tt}}}{2\pi\sqrt{g_{rr}}} ~.
  \eqno{({\rm A}.14)}$$
On the other hand, because $\kappa(r_{h})$ is a constant on the
horizon, this means that, in formula (A.13), the dependence of
$T_{H}$ on the variable $\theta$ is only apparent.

\end{document}